\documentclass[final,1p,times]{elsarticle} 
\usepackage{graphicx} 
\usepackage{amssymb} 
\usepackage{amsthm} 
\usepackage{lineno} 

\usepackage{amsmath,amssymb,amsthm}
\usepackage{graphicx}
\usepackage{bm}
\usepackage{color}

\journal{Nuclear Physics A} 
\begin{document} 

\newcommand{\vAi}{{\cal A}_{i_1\cdots i_n}} 
\newcommand{\vAim}{{\cal A}_{i_1\cdots i_{n-1}}} 
\newcommand{\vAbi}{\bar{\cal A}^{i_1\cdots i_n}}
\newcommand{\vAbim}{\bar{\cal A}^{i_1\cdots i_{n-1}}}
\newcommand{\htS}{\hat{S}} 
\newcommand{\htR}{\hat{R}}
\newcommand{\htB}{\hat{B}} 
\newcommand{\htD}{\hat{D}}
\newcommand{\htV}{\hat{V}} 
\newcommand{\cT}{{\cal T}} 
\newcommand{\cM}{{\cal M}} 
\newcommand{\cMs}{{\cal M}^*}
\newcommand{\vk}{\vec{\mathbf{k}}}
\newcommand{\bk}{\bm{k}}
\newcommand{\kt}{\bm{k}_\perp}
\newcommand{\kp}{k_\perp}
\newcommand{\km}{k_\mathrm{max}}
\newcommand{\vl}{\vec{\mathbf{l}}}
\newcommand{\bl}{\bm{l}}
\newcommand{\bK}{\bm{K}} 
\newcommand{\bb}{\bm{b}} 
\newcommand{\qm}{q_\mathrm{max}}
\newcommand{\vp}{\vec{\mathbf{p}}}
\newcommand{\bp}{\bm{p}} 
\newcommand{\vq}{\vec{\mathbf{q}}}
\newcommand{\bq}{\bm{q}} 
\newcommand{\qt}{\bm{q}_\perp}
\newcommand{\qp}{q_\perp}
\newcommand{\bQ}{\bm{Q}}
\newcommand{\vx}{\vec{\mathbf{x}}}
\newcommand{\bx}{\bm{x}}
\newcommand{\tr}{{{\rm Tr\,}}} 
\newcommand{\bc}{\textcolor{blue}}

\newcommand{\beq}{\begin{equation}}
\newcommand{\eeq}[1]{\label{#1} \end{equation}} 
\newcommand{\ee}{\end{equation}}
\newcommand{\bea}{\begin{eqnarray}} 
\newcommand{\eea}{\end{eqnarray}}
\newcommand{\beqar}{\begin{eqnarray}} 
\newcommand{\eeqar}[1]{\label{#1}\end{eqnarray}}
 
\newcommand{\half}{{\textstyle\frac{1}{2}}} 
\newcommand{\ben}{\begin{enumerate}} 
\newcommand{\een}{\end{enumerate}}
\newcommand{\bit}{\begin{itemize}} 
\newcommand{\ec}{\end{center}}
\newcommand{\bra}[1]{\langle {#1}|}
\newcommand{\ket}[1]{|{#1}\rangle}
\newcommand{\norm}[2]{\langle{#1}|{#2}\rangle}
\newcommand{\brac}[3]{\langle{#1}|{#2}|{#3}\rangle} 
\newcommand{\hilb}{{\cal H}} 
\newcommand{\pleft}{\stackrel{\leftarrow}{\partial}}
\newcommand{\pright}{\stackrel{\rightarrow}{\partial}}

\begin{frontmatter} 


\title{Dynamical Effects on Jet Energy Loss in a QCD Medium}

\author{Magdalena Djordjevic}

\address{Department of Chemistry and Physics, Arkansas State University,
State University, AR, 72467, USA}

\begin{abstract} 
Computation of radiative energy loss in a finite size dynamically screened QCD medium is a key
ingredient for obtaining reliable predictions for jet quenching in ultra-relativistic heavy ion collisions.
We develop a theory which allows calculating, to first order in the number of scattering centers, the energy loss of a heavy quark traveling through a finite size dynamical QCD medium. We show that the result for a dynamical medium is significantly larger compared to a medium consisting of randomly distributed static scattering centers. Therefore, a quantitative description of jet suppression at RHIC and LHC experiments must correctly account for the dynamics of the medium's constituents. Furthermore, qualitative predictions that come from this energy loss formalism are also presented.
\end{abstract} 

\end{frontmatter} 



\section{Introduction}
Heavy flavor suppression is considered to be a powerful tool to study the properties of a QCD medium created in ultra-relativistic heavy ion collisions~\cite{Brambilla}. The suppression results from the energy loss of high energy partons moving through the plasma~\cite{suppresion}. Therefore, 
reliable computations of heavy quark energy loss mechanisms are
essential for the reliable predictions of jet suppression.

In majority of currently available studies the medium-induced radiative 
energy loss is computed by assuming that the QCD medium consists of randomly 
distributed  {\em static} scattering centers (``static QCD medium''). However, 
in reality, constituents of the medium are dynamical, and we recently showed 
that inclusion of dynamical QCD medium effects are important in the radiative 
energy loss calculations~\cite{DH_Inf,DH_PRL}. Furthermore, calculation of the 
energy loss has to be performed in a finite size QCD medium, since the size of 
the QCD medium created in both RHIC and LHC is in reality finite. While 
methods for energy loss calculation have been developed for an infinite optically 
thick dynamical QCD medium, no such approach 
exists for finite optically thin medium. However, it is of crucial importance to develop the energy loss formalism for the case of a finite size optically thin medium, in order to make reliable predictions applicable for the range of parameters relevant for RHIC  and LHC experiments.

In this paper we concentrate on the theory of heavy quark radiative energy loss in a {\em finite size dynamical} QCD medium, and present some qualitative theoretical predictions relevant for the upcoming LHC experiments. Only the main results for heavy quark energy loss are presented here. For more detailed version see~\cite{DH_Inf,DH_PRL,MD_formalism}, and references therein.
\section{Heavy quark energy loss}
We compute the medium-induced radiative energy loss for a heavy quark to the first (lowest) order in number of scattering centers. To calculate this process, we consider the radiation of one gluon induced by one collisional interaction with the medium. In distinction to the static case, we take into account that the collisional interactions are exhibited with dynamical (moving) medium partons. We consider a finite size QCD medium, and to model this medium we adopt the approach from Zakharov~\cite{Zakharov}. That is, we assume that medium has a size L, and that the collisional interaction has to occur inside the medium. The calculations were performed by using two Hard-Thermal Loop approach, and the explicit calculation \cite{MD_formalism} of all diagrams contributing to 
first order in the opacity yields the following expression for the
fractional radiative energy loss ($\mu$ is the
Debye screening mass and parametrizes the density of the medium):
\beqar
\frac{\Delta E_{\mathrm{dyn}}}{E} 
&=& \frac{C_R \alpha_s}{\pi}\,\frac{L}{\lambda_\mathrm{dyn}}  
    \int dx \,\frac{d^2k}{\pi} \,\frac{d^2q}{\pi} \, 
    \frac{\mu^2}{\bq^2 (\bq^2{+}\mu^2)}\,
\nonumber \\
&& \, \, \times
    \left(1-\frac{\sin({\frac{(\bk{+}\bq)^2+\chi}{x E^+} \, L})} 
    {\frac{(\bk{+}\bq)^2+\chi}{x E^+}\, L}\right) 
    \frac{2(\bk{+}\bq)}{(\bk{+}\bq)^2{+}\chi}\cdot
    \left(\frac{(\bk{+}\bq)}{(\bk{+}\bq)^2{+}\chi}
    - \frac{\bk}{\bk^2{+}\chi}
    \right) \!.
\eeqar{DeltaEDyn}
Here $\bk$ and $\bq$ are the transverse momenta of radiated and exchanged gluon respectively,  $\lambda_\mathrm{dyn}^{-1} \equiv 3 \alpha_s T$ 
defines the ``dynamical mean free path'' \cite{DH_Inf}, $\alpha_s$ is the strong coupling constant, and $C_R{=}\frac{4}{3}$. Further, 
$\chi\equiv M^2 x^2 + m_g^2$, where $M$ is the heavy quark mass, 
$x$ is the longitudinal momentum 
fraction of the heavy quark carried away by the emitted gluon and 
$m_g^2=\frac{1}{2}\mu^2$ is the effective mass squared for gluons with 
hard momenta $k\gtrsim T$.

This result allows us to 
compare the radiative energy loss in dynamical and static~\cite{DG_Ind} QCD media, from which one observes that the expressions are remarkably similar, with two important differences~\cite{DH_PRL,MD_formalism}. First, the effective mean free paths are different, and this difference increases the energy loss in the dynamical medium by ~20$\%$. The second difference is the change in the effective differential cross sections, which is reflected by the following expression
  $\left[\frac{\mu^2}{(\bq^2{+}\mu^2)^2}\right]_\mathrm{stat}
  \mapsto
  \left[ \frac{\mu^2}{\bq^2 (\bq^2{+}\mu^2)} \right]_\mathrm{dyn}$, which
significantly increases the energy loss rate, as we can see in the following 
Section. 
\section{Numerical results}
In this section we present some numerical results for the radiative energy loss. The left panel of the Fig.~\ref{RHICLHC} shows the ratio of the radiative energy loss in dynamical and static QCD media as a function of jet energy at RHIC. The ratio is shown by the solid curve in the figure. Note that the value of 1 on the vertical axis would correspond to the case where there is no difference between the energy loss in dynamical and static QCD media. The dark gray region corresponds to the 20$\%$ increase in the energy loss due to the difference in the effective "path lengths". The 
light gray region corresponds to additional 40$\%$ increase in the energy loss due to the difference in effective differential crossections. Therefore, we can conclude that, at RHIC, there is no jet energy domain where the assumption of static scatterers is valid. This conclusion applies to all types of quarks, both light and heavy, as shown in~\cite{DH_PRL}.
\begin{figure}[ht]
\centering
\includegraphics[scale=0.6]{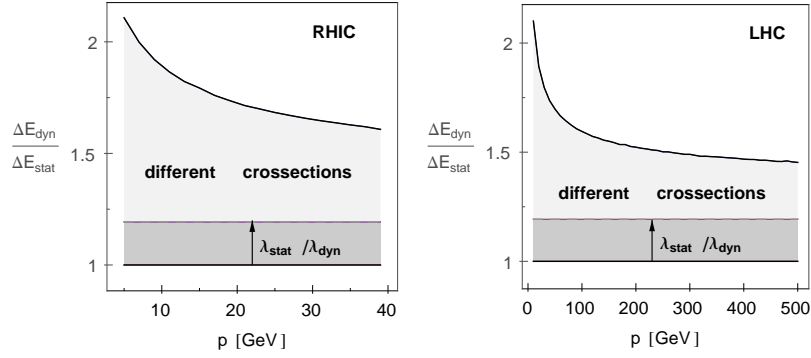}	       
\caption[]{Ratio of the fractional radiative energy loss in finite size dynamical and static QCD media for charm quark as a function of initial quark energy E (solid curve).
Left (right) panel corresponds to RHIC (LHC) conditions. For parameter values see~\cite{DH_PRL}.}
\label{RHICLHC}
\end{figure}

A similar conclusion is valid for the LHC case as well, as can be seen in the right panel of Fig.~\ref{RHICLHC}. However, from this figure we also notice that, in a finite size QCD medium, the difference between the energy losses in dynamical and static QCD media decreases with the increase in jet energy. The decrease in ratio is even more transparent if we asymptotically increase the jet energy. That is, at asymptotic limit, a static QCD medium becomes valid approximation. The reason behind this is that, at asymptotically large jet energies, we recover the LPM limit of the energy loss. The ratio of the radiative energy losses in finite size dynamical and static QCD media then becomes
 $ \lim_{E \rightarrow \infty} \frac{\Delta E_{\mathrm{dyn}}}
  {\Delta E_{\mathrm{stat}}}= \lim_{E \rightarrow \infty}
  \frac{\lambda_\mathrm{stat}} {\lambda_\mathrm{dyn}} 
  \frac{\ln\frac{4 E T}{\mu^2}}{\ln\frac{4 E T}{\mu^2}{-}1} 
  = \frac{\lambda_\mathrm{stat}} {\lambda_\mathrm{dyn}}$~\cite{DH_PRL,MD_formalism}. This leads
to the conclusion that, at asymptotically large jet energies, the static approximation becomes valid, up to a multiplicative constant which can be renormalized.

\begin{figure}[ht]
\centering
\includegraphics[scale=0.65]{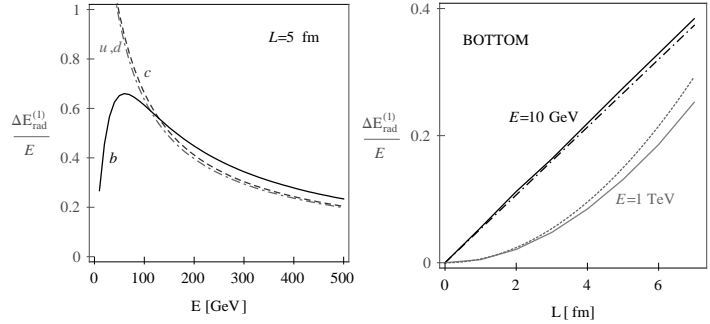}	       
\caption[]{{\em Left panel}: Fractional $1^{st}$ order in opacity radiative energy loss for an assumed 
path length L = 5 fm as a function of jet energy for light, charm and
bottom quarks (dot-dashed, dashed and solid curve, respectively). 
{\em Right panel}: Fractional $1^{st}$ order in opacity radiative energy loss as a function of thickness of the medium for bottom quark. Black curves correspond to initial jet energy of E = 10GeV,
while gray curves correspond to E =1TeV. Black dot-dashed curve shows what 
would be the fractional energy loss in infinite size QCD medium for 10GeV jet, 
while dotted gray curve shows what would be the fractional energy loss in LPM limit for 1 TeV jet. Note that for each panel, we assume a medium of
temperature T = 400 MeV (``LHC conditions'').}
\label{LHC_EB}
\end{figure}

Finally, we discuss some qualitative conclusions arising from this newly developed energy loss formalism. To do this, we concentrate on the LHC case, and the left panel of Fig.~\ref{LHC_EB} shows the fractional energy loss as a function of initial jet energy. We see that, in the range up to 70 GeV, light and charm quark lose all of their energy, so we can not distinguish them in that range. On the other hand, in the range above 70 GeV, they loose the same amount of energy, so they can not be distinguished either. Therefore, one expects that at LHC, charm and light quarks will have similar suppressions, leading to a conclusion that charm suppression is not a good probe for studying the "dead-cone effect"~\cite{DG_Ind,dead_cone}. On the other hand, we see that bottom energy loss is significantly different from charm and light energy loss. This therefore leads to a conclusion that bottom suppression is an excellent probe to study the finite mass effect on the energy loss at LHC. Note that, despite the fact that charm is not a good probe for studying the "dead-cone effect", the fact that charm presents a clear independent probe still makes charm an excellent probe for testing our understanding of the QCD matter created at LHC. 

Furthermore, another interesting observation can be obtained from the bottom energy loss dependence on the thickness of the medium, which is shown in the right panel of Fig.~\ref{LHC_EB}. For 10 GeV jet, we see that jet energy loss thickness dependence almost ideally overlaps with the incoherent Berch-Gunion limit, while for 1TeV jet energy loss thickness dependence almost ideally overlaps with LPM limit. Therefore, we see that, at the jet energy range applicable to LHC, we can observe a perfect transition of the bottom energy loss thickness dependence, from incoherent linear to the coherent quadratic dependence characteristic for the LPM effect. 
\section{Summary}
We have studied the radiative energy loss for both light and heavy quarks, in a finite size dynamical QCD medium. We showed that the energy loss in a dynamical medium is significantly larger than in a static medium for the entire range of experimentally realistic energies. Therefore, the constituents of a QCD medium can not be approximated as static scattering centers. i.e. the dynamical effects have to be included for reliable predictions of jet suppression in the upcoming experiments.
The formalism suggests new qualitative, experimentally observable phenomena. Detailed quantitative predictions relevant for RHIC and LHC will follow.



\begin{thebibliography}{00} 
 

\bibitem{Brambilla} N. Brambilla et al., Preprint hep-ph/0412158 (2004).

\bibitem{suppresion} M. Guylassy, I. Vitev, X. N. Wang and B. W. Zhang, in 
Quark Gluon Plasma 3, edited by R. C. Hwa and X. N. Wang, p. 123 
(World Scientific, Singapore, 2003)  

\bibitem{DH_Inf}
  M. Djordjevic and U. Heinz, submitted to Phys. Rev. C (2007)  
 
\bibitem{DH_PRL} M. Djordjevic and U. Heinz, Phys. Rev. Lett. {\bf 101}, 
022302 (2008).

\bibitem{MD_formalism} M. Djordjevic, Preprint arXiv:0903.4591 [nucl-th] (2009). 
\bibitem{Zakharov} B.~G.~Zakharov, JETP Lett. {\bf 76}, 201 (2002).

\bibitem{DG_Ind}
  M.~Djordjevic and M.~Gyulassy, Phys.\ Lett.\ B \ {\bf 560}, 37 (2003); 
  and Nucl.\ Phys.\ A {\bf 733}, 265 (2004).

\bibitem{dead_cone} Y. L. Dokshitzer and D. E. Kharzeev, 
Phys. Lett. B {\bf 519}, 199 (2001).

\end{thebibliography}
\end{document}